\documentclass[12pt,a4paper]{article}
\usepackage{amsmath,amssymb}
\usepackage{graphicx,epsfig}
\usepackage{color}

\begin{document}
\baselineskip 0.6cm

\begin{titlepage}

\begin{flushright}
\end{flushright}

\vskip 1.0cm

\begin{center}
  {\Large \bf The Higgs mass as a function  \\
    of the compactification scale}
 
  \vskip 1.0cm
 
  {\large Riccardo Barbieri, Guido Marandella and Michele Papucci}
 
  \vskip 0.5cm
 
  {\it Scuola Normale Superiore and INFN, Piazza dei Cavalieri 7,
    I-56126 Pisa, Italy} \\
 
  \vskip 1.0cm
 
  \abstract{We calculate to a few percent precision the Higgs
    potential in a model with supersymmetry broken by boundary
    conditions on an extra-dimension, compactified to a segment of
    length $L$, and a top quark quasi-localized on one of the two
    boundaries. $1/L$ alone, in the range 2-4 TeV, determines the
    Higgs mass, in the range 110-125 GeV, and the spectrum of
    gauginos, higgsinos and of the third-generation squarks. Lower
    values of $1/L$ cannot be excluded, with a progressive
    delocalization of the top quark.}

\end{center}
\end{titlepage}

\section{Introduction}
\label{sec:intro}

In the Standard Model (SM) the sensitivity of the Fermi scale
$G_F^{-1/2}$ to the cut-off $\Lambda$ strongly suggests the presence
of new physics close to $G_F^{-1/2}$ itself. Since long time the
possibility is contemplated that such new physics be represented by
supersymmetric particles with masses of order $G_F^{-1/2}$. Relatively
more recently, the scale of a compactified extra dimension (one or
more) has also been put forward to play the same role
\cite{Antoniadis}. In this paper we study, as precisely as possible,
the consequences for ElectroWeak Symmetry Breaking (EWSB) from the
merging of these two ideas in a definite scheme, as defined in
\cite{Barbieri:2002sw} \footnote{See also
  \cite{Marti:2002ar,Barbieri:2002uk}}.

This contamination has a main motivation: it allows to describe
supersymmetry breaking in terms of a single parameter, $L$, the length
of the segment to which the extra-dimension is compactified. Although
$L \rightarrow \infty$ is the limit where supersymmetry is recovered,
the compactification scale is not the scale above which a
supersymmetric spectrum approximately appears, due to the presence of
the Kaluza-Klein (KK) modes. There is no such scale. In fact, the KK
modes play a crucial role in rendering more precise the supersymmetric
softening of the ultraviolet divergences. The introduction of the
usual soft supersymmetry-breaking parameters is avoided. Nor there is
any need of introducing a supersymmetric $\mu$-parameter, since only
one Higgs field gets a non-vanishing vacuum expectation value
consistently with all phenomenological requirements and with the
supersymmetry constraints \cite{Barbieri:2000vh}.

For a theory that is designed to render the Fermi scale insensitive to
the cut-off, the connection between $G_F^{-1/2}$ and the scale of new
physics should be determined as neatly as possible. This is highly
desirable not only theoretically but also from a phenomenological and
pragmatic point of view, since the possibility to test the theory at
the Tevatron or at the LHC crucially depends on this property. This
motivates a careful study of the Higgs potential. A peculiar feature
that emerges from this study is the gap that can result between $1/L$
and $G_F^{-1/2}$ or, even more so, the Higgs mass. Although
anticipated in \cite{Barbieri:2002sw}, the precise assessment of this
property as the equally precise determination of the Higgs mass as a
function of $1/L$ -- our main goals in this paper -- require the two
loop calculation of the Higgs potential described below.

The paper is organized as follows. The model is defined in Sect.
\ref{sec:model}. In Sect. \ref{sec:potential} the Higgs potential is
calculated to two loops accuracy in $\alpha_s=g_s^2 /(4\pi)$ and
$\alpha_t=y_t^2/(4\pi)$ with the top supermultiplet exactly localized
at $y=0$. In Sect.  \ref{sec:quasiloc-top} a quasi localized top is
introduced and both the Higgs mass and the determination of the Fermi
scale are studied as function of $1/L=2 \div 4 $ TeV. The issue of the
uncertainties is explicitly addressed in Sect.
\ref{sec:uncertainties}. In Sect.  \ref{sec:low-1overL} the
possibility is considered of a lower value of the compactification
scale, allowing for the presence of a common mass for the Higgs
hypermultiplets. The spectrum of the model is given in Sect.
\ref{sec:spectrum}. The summary and the conclusions are drawn in Sect
\ref{sec:conclusions}.

\section{The Model}
\label{sec:model}

We consider a 5D, $SU(3) \times SU(2) \times U(1)$ invariant theory
with every multiplet of the SM, gauge, matter or Higgs, promoted to a
$N=1$ supermultiplet (hypermultiplet). The fifth dimension is meant to
be compactified on a segment parameterized by $y \in \left[ 0,L
\right]$ \footnote{In Refs.
  \cite{Barbieri:2002sw,Barbieri:2002uk,Barbieri:2000vh} $R=2 L/ \pi$
  has been used.}.

As advocated long ago by Scherk and Schwarz \cite{Scherk:1978ta}, we
break supersymmetry by boundary conditions at $y=0,L$ that distinguish
the different fields in every hypermultiplet. Given the 5D gauge
theory, there is a single choice of these boundary conditions,
consistent with the symmetries of the theory and with supergravity,
that leads to a massless spectrum identical to the SM one
\cite{Barbieri:2000vh}. This non trivial property is at variance with
what happens when global supersymmetry is spontaneously broken in a 4D
theory, where a hidden sector and a mediation mechanism of
supersymmetry breaking have to be introduced. With these boundary
conditions, summarized in Table \ref{tab:kktower}, every field
acquires, in the effective 4D picture, a definite KK tower. The first
KK states of the SM particles are at $\pi/L$, whereas the extra
particles implied by supersymmetry and/or Poincar\'e invariance are at
$\pi/(2 L)$ or higher.

\begin{table}[!bt]
  \centering
    \begin{tabular}{|c|c|c|c|}
\hline
\hline
      $\psi_M$, $\phi_H$, $A^{\mu}$ &  $\phi_M$, $\psi_H$, $\lambda$ &
      $\phi^c_M$, $\psi^c_H$, $\lambda^c$ &  $\psi^c_M$,
      $\phi^c_H$, $\phi_{\Sigma}$ \\
\hline
    $ (+,+) $ & $ (+,-) $ & $ (-,+) $ & $ (-,-) $ \\
\hline
\hline
    \end{tabular}

  \caption{Boundary conditions for
   gauge $(A_{\mu},\lambda,\lambda^c,\phi_{\Sigma})$, matter $(\psi_M,
   \phi_M,\psi_M^c,\phi_M^c)$ and Higgs
   $(\phi_H,\psi_H,\phi_H^c,\psi_H^c)$ multiplets at $y=0$ and $y=L$.}
  \label{tab:kktower}
\end{table}

A theory thus defined has a divergent Fayet-Iliopoulos term induced at
one loop \cite{Ghilencea:2001bw}, which can be canceled by introducing
a second Higgs-like hypermultiplet, $H_c$. In this case one obtains,
at tree level, two (massless) Higgs-like scalars as in the Minimal Supersymmetric
Standard Model (MSSM). At variance with the MSSM, however, the second
Higgs-like scalar does not get a vacuum expectation value (VEV) and
plays no role in EWSB.

Without affecting the massless spectrum, it is consistent to introduce
suitable supersymmetric mass terms $M$ for the matter hypermultiplets
\cite{Barbieri:2002ic}. They deform the massive spectrum as well as
the wave function in $y$ of every state \cite{Barbieri:2002uk}. The
wave function of the massless states (the matter fermions) becomes an
exponential, $\psi_0 (y) \propto \left| M \right|^{1/2} e^{-M y + (M-
  \left| M \right|)L/2}$, so that in the large $ML$ limit, $| ML | \gg
1$, the massless fermion gets localized on one of the two boundaries,
$y=0$ or $y=L$, depending on the sign of $M$. At the same time a
chiral $N=1$, $4D$ supermultiplet is recovered with one scalar
becoming also massless and localized. All the other states get a heavy
mass, increasing like $M$. These hypermultiplet masses, in general
different for every matter multiplet, are non renormalized parameters
and could be fixed in a more fundamental theory. Not to introduce
again a Fayet-Iliopoulos term they have to satisfy simple conditions
\cite{Barbieri:2002ic}: e.g. a common mass for the quark
hypermultiplets of a given generation, $M_{Q_i}$, for the lepton
hypermultiplets, $M_{L_i}$, and for the Higgs-like hypermultiplets
$M_H$. We stick to this configuration in the following.

The top Yukawa coupling, necessarily localized at one of the two
boundaries, say $y=0$, is the source of EWSB, which is therefore
influenced by the mass terms of the top quark hypermultiplets $Q$ and
$U$, $M_{Q_3} \equiv M$. We shall take them quite closely localized at
$y=0$, with $ML \gtrsim 1$ for most of the paper. For this reason, it
is a significant approximation to consider first what happens when $Q$
and $U$ are exactly localized.

\section{The Higgs potential with a top localized on one boundary}
\label{sec:potential}

Symbolically, the reference Lagrangian is
\begin{equation}
\label{eq:lagrangian5D}
\mathcal{L}=\mathcal{L}_{5D}^{N=1} \left(\textrm{gauge} \oplus 
\textrm{Higgs} \right)+\mathcal{L}_{4D}^{N=1} \left(Q \oplus U 
\right) \delta (y)
\end{equation}
where the $4D$, $N=1$ supersymmetric Lagrangian for the $Q,U$ chiral
supermultiplets includes the top Yukawa coupling

\begin{equation}
\label{eq:lagrangian4D}
\mathcal{L}_{4D}^{N=1}  \left(Q \oplus U \right) = \int \textrm{d}^4 \theta 
Q^{\dagger} e^V Q + \int \textrm{d}^4 \theta U^{\dagger} e^V U+ 
\left[ \int \textrm{d}^2 \theta \lambda_t H Q U + \textrm{h.c.} 
\right]
\end{equation}
We ignore for the time being possible kinetic terms for the gauge and
Higgs multiplets also localized at the boundaries (See Sect.
\ref{sec:uncertainties}).

With the boundary conditions in Table \ref{tab:kktower}, after the
integration over $y$, the tree level potential for the real part of
the zero mode of the Higgs field is

\begin{equation}
\label{eq:vtree}
V^{\textrm{tree}}(h^2)= \frac{g^2+g^{\prime 2}}{32} h^4
\end{equation}

We are interested in the effective potential $V(h^2)$ which, expanded
around $h=v$, gives

\begin{equation}
\label{eq:veffective}
V(h^2) \simeq 2 v \; V' (v^2) \; h + \left[ V'(v^2)+2 v^2 \; V''(v^2) 
\right] \; h^2
\end{equation}
Hence
\begin{equation}
\label{eq:minimization}
V' (v^2)=0
\end{equation}
is the equation which determines $v$ or the Fermi scale $G_F^{-1/2}$ and
\begin{equation}
\label{eq:higgsmass}
m_h^2= 4v^2 \; V''(v^2)
\end{equation}
is the physical Higgs squared mass. We aim to an accuracy of a few
percent in $V'$ and $V''$.

The 1 loop electroweak contribution to $V'$ has been computed in
\cite{Antoniadis:1998sd}
\begin{equation}
\label{eq:1loopgauge}
\delta V'_{\textrm{ew}}(v^2) \simeq \delta V'_{\textrm{ew}}(0)= 
\frac{7 \zeta(3) (3 g^2+g'^2)}{128 \pi^2 L^2} = 0.93 
\frac{10^{-2}}{L^2}
\end{equation}
up to corrections of relative order $(gvL)^2$.

The one loop $(g^2+g'^2) \alpha_t$ correction to $V''$, in localized
approximation for the top, coincides, at logarithmic level, with the
same correction in the MSSM for appropriate values of the stop masses
$m^2_Q$, $m_U^2$ (see below). Its contribution to the Higgs squared
mass is \cite{Brignole:1992uf}
\begin{equation}
\label{eq:1loop-alfatop-alfa}
\delta m_h^2 \left( (g^2+g'^2) \alpha_t \right)= 
\frac{-3}{4\,{\sqrt{2}}\,{\pi }^2}{G_F} \; {{m_t}}^2 \; {{M_Z}}^2 \; 
\log \frac{m^2_Q \; m_U^2}{m_t^4}
\end{equation}

Our task is then to compute to the relevant order of approximation the
$(\alpha_t,\alpha_s)$-dependent contributions to eqs.
(\ref{eq:minimization}) and (\ref{eq:higgsmass}).

\subsection{$(\alpha_t,\alpha_s)$-corrections. General expressions}
\label{sec:alfatop-alfastrong-express}

The corrections of interest contain log's of the fine structure
constants, $\alpha_s$ and $\alpha_t$, which arise from the infrared
behavior of the integrals.  This is due to the masslessness of the
squarks $\widetilde Q$ and $\widetilde U$ at tree level, which become
massive only at one loop. To deal properly with this situation we
introduce, as infrared regulators, the squark masses $m^2_{0,Q}$ and
$m^2_{0,U}$ for the two multiplets, also localized at
$y=0$\footnote{This also helps in keeping right track of the order in
  the loop expansion.}. These masses will be sent to zero at the end
of the calculation.  With these masses the potential of interest,
$\delta V_{\textrm{top}} (v^2)$, has a one loop contribution

\begin{align}
\label{eq:1loop-top-localized}
\delta V_{\textrm{top}}^{\textrm{1 loop}} = 3 \int \frac{d^4 p}{(2
  \pi)^4} \left[ \right. & \log\left(
  p^2+m^2_{0,Q}+m_{0,t}^2\right)+\log\left(
  p^2+m^2_{0,U}+m_{0,t}^2\right) \nonumber \\
& -2\log\left( p^2+m_{0,t}^2\right) \left. \right]
\end{align}
where $m_{0,t}=y_t v/ \sqrt{2}$ is the unrenormalized top quark mass,
and a two loop contribution $\delta V_{\textrm{top}}^{\textrm{2
    loop}}$ which arises from the diagrams in Fig.
\ref{fig:2ndtopology}, in superfield notation, and is explicitly given
in Appendix \ref{app:2loop}.

\begin{figure}[t]
  \centering \includegraphics[width=10cm]{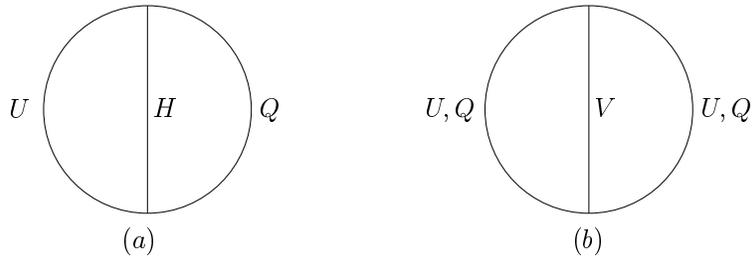}
\caption{The diagrams that contribute to the Higgs potential at order
  $\alpha_s \alpha_t$ and $\alpha_t^2$ in superfield
  notation.}\label{fig:2ndtopology}
\end{figure}

The propagators for all the components of the $Q,U$ supermultiplets
are in the background of the field $h$. Since $Q$ and $U$ propagate in
ordinary Minkowsky space, at $y=0$, the only components of the Higgs
and gauge supermultiplets, $H$ and $V$, that contribute in Fig.
\ref{fig:2ndtopology} are those with (+) boundary conditions at $y=0$.
Up to trivial kinematic factors, after the Wick rotation, their
propagators are proportional to
\begin{align}
  S^{+,+}(k) &\propto \coth \sqrt{k^2}L \nonumber\\
  S^{+,-}(k) &\propto \tanh \sqrt{k^2}L. \label{eq:propagators}
\end{align}
From
\begin{equation}
\delta V_{\textrm{top}}=\delta V_{\textrm{top}}^{\textrm{1 
loop}}+\delta V_{\textrm{top}}^{\textrm{2 loop}}
\end{equation}
the contributions to $v^2 \delta V' (v^2)$ and $v^4 \delta V'' (v^2)$
are finite after reexpressing $m_{0,t},m^2_{0,Q},m^2_{0,U}$  in (\ref{eq:1loop-top-localized}) in terms of the physical masses\footnote{$m_Q$
  is the mass of the stop-left. For the mass of the sbottom-left,
  which only enters in the two loop diagrams, we use
  $m_B^2=m_Q^2-m_t^2$.}
\begin{subequations}
\label{eq:renormalizations}
\begin{align}
  m_t^2 &= m_{0,t}^2 \left(1+B_{\psi}^U +B_{\psi}^Q +2
    Z_{y_t}\right) \label{eq:renormalization-conditions-top} \\
  m^2_Q &= m^2_{0,Q}
  (1+B_{\varphi}^Q)+m_{0,t}^2(1+B_{\varphi}^Q+B_F^U)+\delta m^2_Q  \label{eq:renormalization-conditions-stopq}\\
  m_U^2 &= m^2_{0,U} (1+B_{\varphi}^U)+m_{0,t}^2(1+B_{\varphi}^U+B_F^Q
  )+\delta m_U^2 \label{eq:renormalization-conditions-stopu}
\end{align}
\end{subequations}
and after making an expansion in the one loop quantities: $\delta
m_{U,Q}^2$, the one loop corrections to the squared masses, $Z_{y_t}$,
the Higgs-top-top vertex correction, and the various $B$-factors, the
corrections to the wave functions of the various fields. To the
precision of interest all these quantities are computed at zero
external momenta, except for those involved in
(\ref{eq:renormalization-conditions-top}) where we use for $m_t$ the
running mass at $p^2=-m_t^2$. The explicit expressions for all these
factors are given in App. \ref{app:b-factors}.

Finally, as anticipated, the fictitious bare masses $m^2_{0,Q}$,
$m^2_{0,U}$ are set to zero. Note that they only appear in $m_Q^2$,
$m_U^2$ in eqs.
(\ref{eq:renormalization-conditions-top})-(\ref{eq:renormalization-conditions-stopq})-(\ref{eq:renormalization-conditions-stopu}).
To leading order in $\alpha_t$ and $\alpha_s$ one has
\cite{Delgado:1998qr}
\begin{subequations} \label{eq:mstops}
\begin{align}
  m_U^2 & = m_t^2 + \frac{7 \zeta(3)}{24 \pi} \frac{8 \alpha_s +6
    \alpha_t}{L^2} + \mathcal{O} (\alpha_t^2, \alpha_t \alpha_s)
\label{eq:mstopu} \\
m^2_Q & = m_t^2 + \frac{7 \zeta(3)}{24 \pi} \frac{8 \alpha_s +3
  \alpha_t}{L^2} + \mathcal{O} (\alpha_t^2, \alpha_t \alpha_s)
   \label{eq:mstopq}
\end{align}
\end{subequations}

\subsection{$(\alpha_t,\alpha_s)$-corrections. Results}
\label{sec:alfatop-alfastrong-results}

To calculate explicitly the corrections of interest we make a
systematic expansion of $v^2 \delta V'_{\textrm{top}} (v^2)$ and $v^4
\delta V''_{\textrm{top}} (v^2)$ in $\alpha_t$, $\alpha_s$ and
$m_Q^2$, $m_U^2$, $m_t^2$, all formally treated as quantities of the
same order. In so doing, care must be taken in avoiding spurious
infrared divergences. In $v^2 \delta V'_{\textrm{top}} (v^2)$ we keep
terms quadratic in these quantities, whereas in $v^4 \delta
V''_{\textrm{top}}$, which starts quadratic in $m_t^2$, we keep those
cubic terms which also include at least a factor of $\log m_Q^2 L^2$,
$\log m_U^2 L^2$ or $\log m_{t}^2 L^2$. In $v^2 \delta
V'_{\textrm{top}} (v^2)$ the $5D$ propagators in
eqs.(\ref{eq:propagators}) are crucial in giving a finite result,
whereas in $v^4 \delta V''_{\textrm{top}} (v^2)$, which is more
convergent in the ultraviolet (but less convergent in the infrared),
the $5D$ propagators can be approximated with their low momentum
expansion. We find
\begin{align}
\label{eq:2loop-quadratic}
v^2 \delta V'_{\textrm{top}} (v^2; m_Q, m_U) = \frac{3 m_t^2}{16
  \pi^2} \, \left\{ \right.  \, & {m_Q}^2\,\left[2\log \left(m_Q
    L\right) - c\right]+ \,{m_U}^2\,\left[2\log \left(m_U L \right)-c
\right] \nonumber \\ - & 2\,{m_t}^2\,\left[2\log \left(m_t L \right)
  -c \right] \left. \right\}
\end{align}
\begin{align}
\label{eq:2loop-quartic}
& v^4 \delta V''_{\textrm{top}} (v^2; m_Q, m_U) = \frac{3 m_t^4}{8
  \pi^2} \log
\left(\frac{m_Q \, m_U}{m_t^2}\right) + \nonumber \\
& \frac{3 m_t^4}{16 \pi^3} \left\{ \frac{m_t^2 G_F}{\pi \sqrt{2}}
  \left[ 2\,{\log^2 \left(\frac{{m_Q}}{{m_t}}\right)} + \log \left(m_t
      L\right)\, \log \left(\frac{{m_Q}}{{m_U}}\right) + {\log^2
      \left(m_U m_Q L^2 \right)} +
    {\log^2 \left(\frac{{m_U}}{{m_t}}\right)} \right] \right. \nonumber \\
& \hspace{1cm} - \frac{8 \alpha_s}{3} \,\left[ {\log^2
    \left(\frac{{m_Q}}{{m_t}}\right)} - 4\,{\log^2 \left(m_t L
    \right)} - 4\,\log \left(m_t L\right)\, \log
  \left(\frac{{m_Q}\,{m_U}}{{{m_t}}^2}\right) +
  {\log^2 \left(\frac{{m_U}}{{m_t}}\right)} \right] \nonumber \\
& \hspace{1cm} + \frac{m_t^2 G_F}{4 \pi \sqrt{2}} \left[ 10\,\log
  \left(m_Q L\right) + 6\,\log \left(m_t L\right) + 12\,\log \left(m_U
    L \right)
\right] \nonumber \\
& \hspace{1cm} - \frac{16 \alpha_s}{3} \left. \left[ \left(1-6\log
      2\right)\,\log \left(\frac{m_Q \, m_U}{m_t^2}\right) - 3\,\log
    \left(m_t L\right) \right] \right\}
\end{align}
where $c= 4 - 2\,\gamma - (12\,\log 2)/7 + 2 \zeta '(3)/\zeta (3)
\simeq 1.33$.

In view of eqs. (\ref{eq:mstops}), the result for $v^2 \delta
V'_{\textrm{top}} (v^2)$ coincides with the result given in Ref.
\cite{Barbieri:2002sw} in logarithmic approximation. Numerically, for
$m_t^{\textrm{pole}}=174.3 \pm 5.1$ GeV, we find
\begin{equation}
\label{eq:2loop-mass-numerical}
\delta V'_{\textrm{top}} (v^2)= - (0.73 \pm 0.05) \frac{10^{-2}}{L^2}
\end{equation}
at $1/L= 3$ TeV, with a negligible residual dependence on $1/L$ of the
coefficient in the range $1/L= 2 \div 4$ TeV due to $(m_t L)^2$ terms.
Note the near cancellation in $ \delta V' (v^2)$, at the 20\% relative
level, between the electroweak term, eq.  (\ref{eq:1loopgauge}), and
the two loop $(\alpha_t,\alpha_s)$-contribution, eq.
(\ref{eq:2loop-mass-numerical}), with a predominance of the first
positive term. To the extent that this calculation is reliable (see
below), the Higgs potential has a positive curvature at $h=0$, so that
EWSB does not take place with exactly localized $Q,U$ multiplets.

For $\delta V'(v^2)$ no simple connection can be established between
this theory and a suitably defined MSSM, because of the difference in
the way higgsinos get a mass: by a $\mu$-term in the MSSM, by pairing
with conjugate states here. Since $\delta V' (v^2)$ is ultraviolet
sensitive in the MSSM, this makes an essential difference. On the
contrary, the stronger ultraviolet convergence of the momentum
integrals in $ \delta V'' (v^2)$, relative to $ \delta V' (v^2)$,
makes it closer to the analogous quantity in a suitably defined MSSM.
The leading $m_t^4$-term coincides. The same is also true for the next
order terms $m_t^4 G_F m_t^2$ and $m_t^4 \alpha_s$, in leading
$\log^2$ approximation, if one compares eq.  (\ref{eq:2loop-quartic})
at $1/L = m_Q=m_U \equiv M_S$ with the MSSM result for $\tan \beta =
\infty, A_t=0$ and all superpartners at $M_S$ \cite{Hempfling:1993qq}.

\section{The case of a quasi-localized top}
\label{sec:quasiloc-top}

As anticipated in Sect. \ref{sec:model}, with the $Q$, $U$
hypermultiplets not exactly localized, all their components acquire a
KK tower of states with M-dependent masses. Most importantly, for
finite $ML$, this tree level spectrum is not supersymmetric. As a
consequence, already at one loop the Higgs potential receives a non
vanishing contribution from $Q$, $U$ exchanges, $\delta
V_{\textrm{top}}^{\textrm{1loop}} (v^2;ML)$, calculated in
\cite{Barbieri:2002uk}. For $ML \leq 2.5$ the slope of this potential,
of negative sign, dominates over the electroweak contribution in eq.
(\ref{eq:1loopgauge}) and triggers EWSB. For $ML \geq 1.5$, however,
$\delta V_{\textrm{top}}^{\textrm{1loop}}$ is not a sufficiently
accurate description of the top-stop contribution to the Higgs
potential, as we shall see explicitly: in localized approximation, $ML
= \infty$, $\delta V_{\textrm{top}}^{\textrm{1loop}}$ vanishes,
whereas the top-stop contribution at two loop does not, as seen in the
previous Section.

\begin{figure}[htbp]
  \centering \includegraphics[width=13cm]{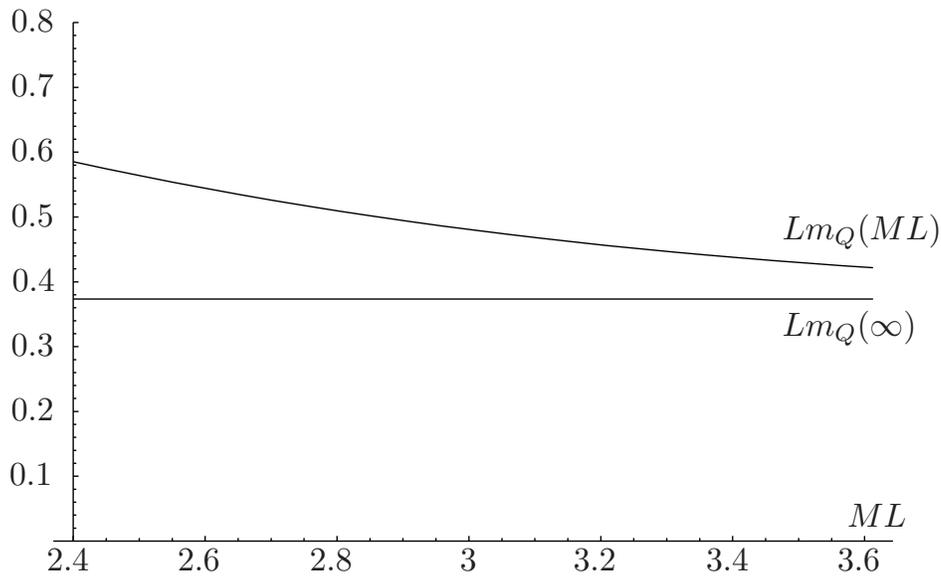}
   \caption{The mass of the left handed stop as a function of 
     the localization parameter $ML$, compared with the same mass in
     the $ML \longrightarrow \infty$ limit.}
   \label{fig:mstop}
\end{figure}

The most important effect of a finite $ML$, compared to $ML = \infty$,
is on the tree level mass of the lightest squarks in the corresponding
KK tower. Although this mass converges exponentially to $m_t$ for the
stops, or to zero for the sbottom left, its effect is still
significant at $ML \simeq 2 \div 3$. In Fig. \ref{fig:mstop} we
compare $m_Q (ML)$ with the corresponding quantity, $m_Q(\infty)=m_Q$,
eq.  (\ref{eq:mstopq}), in localized approximation. The radiative one
loop contribution is only weakly sensitive to $ML$ and dominates over
the tree level mass. Nevertheless the deviation of $m_Q(ML)$ from
$m_Q$ in the region of interest is sizeable. A similar situation holds
for $m_U(ML)$. To account for this effect in the potential, we
consider the first and second derivatives of $\delta V_{\textrm{top}}$
in eqs.  (\ref{eq:2loop-quadratic}) and (\ref{eq:2loop-quartic}) with
$m_Q$ and $m_U$ replaced by $m_Q(ML)$ and $m_U(ML)$ respectively, so
that
\begin{equation}
   \label{eq:potential-replace}
   \delta V_{\textrm{top}} (v^2, ML) \equiv \delta V_{\textrm{top}}(v^2; m_Q(ML),m_U(ML))
\end{equation}
A better approximation of $\delta V_{\textrm{top}}$ (of its
derivatives) is in fact the following
\begin{equation}
   \label{eq:potential-replace2}
   \delta V_{\textrm{top}} (v^2, ML) \equiv \delta V_{\textrm{top}}^{\textrm{1loop}} (v^2, ML)+ 
\delta V_{\textrm{top}}^{\textrm{2loop}} (v^2, ML)
\end{equation}
where
\begin{equation}
   \label{eq:potential-2loop-ML}
    \delta V_{\textrm{top}}^{\textrm{2loop}} (v^2, ML) =  \delta V_{\textrm{top}}(v^2; 
m_Q(ML),m_U(ML))-\delta V_{\textrm{top}}(v^2; m_Q^{\textrm{tree}}(ML),m_U^{\textrm{tree}}(ML))
\end{equation}
properly subtracted to avoid double counting with the one loop term.
For $ML \geq 2$, however, the difference between
(\ref{eq:potential-replace}) and (\ref{eq:potential-replace2}) is
negligible.

\subsection{Determination of the Fermi scale}
\label{sec:Fermi-scale}

\begin{figure}[htbp]
  \centering \includegraphics[width=13cm]{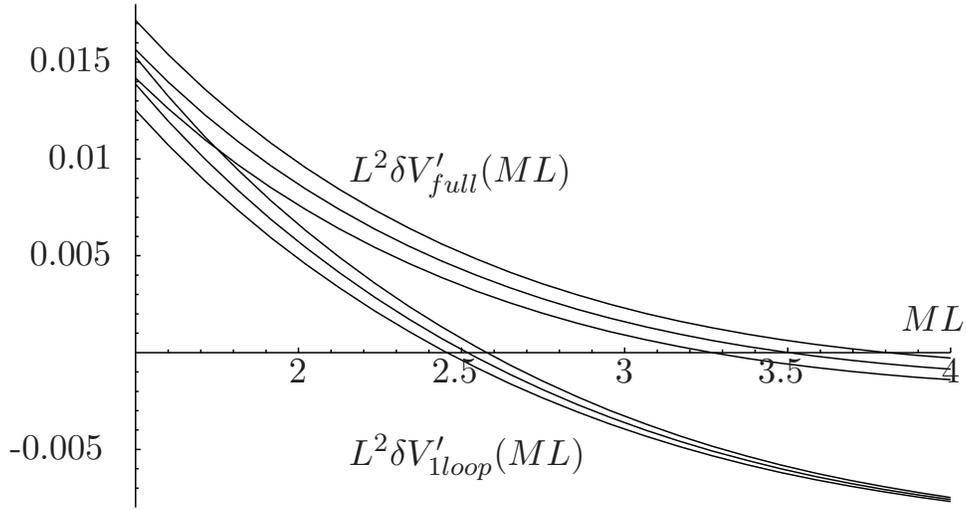}
   \caption{Slope of the full radiative Higgs potential, as discussed 
     in the text, versus the top localization parameter $ML$, compared
     to the one loop approximation, for $m_t^{pole}=174.3 \pm 5.1 {\rm
       \mbox{ }GeV}$.}
   \label{fig:vslope}
\end{figure}

With the inclusion of the tree level contribution from eq.
(\ref{eq:vtree}), the minimum equation (\ref{eq:minimization}) reads
\begin{equation}
   \label{eq:pot-minimum}
   \frac{M_Z^2}{4}=- \delta V'(v^2)
\end{equation}
which must be viewed as a relation between the compactification scale
$1/L$ and the localization parameter $ML$. Fig. \ref{fig:vslope} shows
\begin{equation}
   \label{eq:deltaV-prime}
   -L^2 \delta V'(v^2) = -L^2 \delta V'_{\textrm{ew}} (v^2)- L^2 \delta V'_{\textrm{top}}
    (v^2; ML)
\end{equation}
with the electroweak contribution given in eq. (\ref{eq:1loopgauge})
and the top contribution from eq. (\ref{eq:potential-replace}) or
(\ref{eq:potential-replace2}).
After rescaling by $1/L^2$, $\delta V' (v^2)$ has no significant
residual dependence on $1/L$ in the region of interest, $1/L \geq 2
\textrm{ TeV}$.  For these values of $1/L$, it is $(M_Z L)^2 \simeq
10^{-3}$, so that the one loop approximation to $\delta
V_{\textrm{top}}'$ is clearly inadequate. The flattening of $\delta
V'(v^2)$ at $ML \simeq 3$, due to the partial cancellation between
$\delta V_{\textrm{top}}'$ and $\delta V_{\textrm{ew}}'$, is important
in reducing the tuning between $ML$ and $1/L$.  Also in view of the
uncertainties to be discussed below, this same flattening of $\delta
V' (v^2)$ makes the precise relation between $ML$ and $1/L$ uncertain.
This has little influence, however, on the relation between the Higgs
mass and $1/L$, as we discuss shortly.

\subsection{The Higgs mass as a function of $1/L$}
\label{sec:Higgs-vs-1overL}

\begin{figure}[!tb]
  \centering \includegraphics[width=13cm]{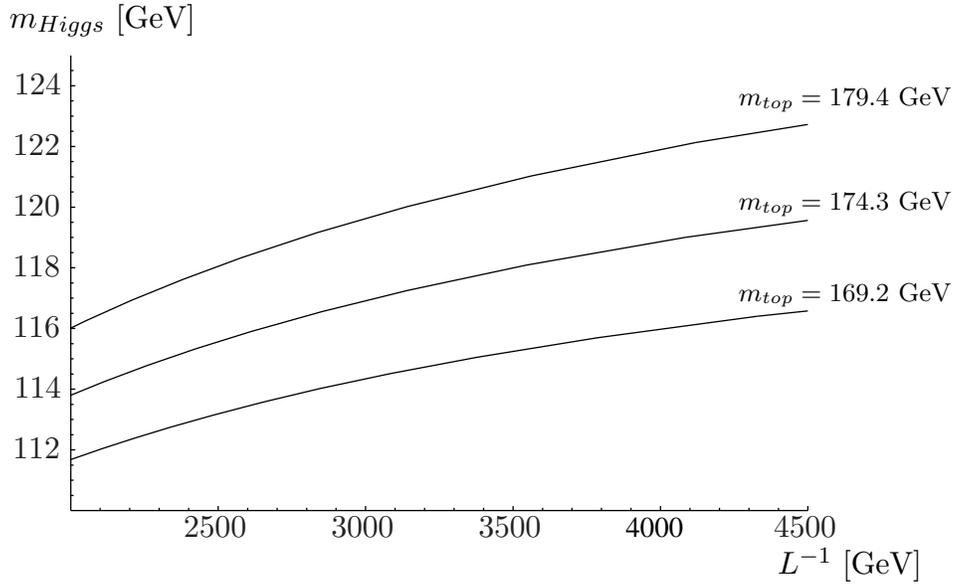}
   \caption{Higgs mass as function of $1/L$ for $m_t^{pole}=174.3 \pm 
     5.1 {\rm \mbox{ }GeV}$.}
   \label{fig:mhiggs}
\end{figure}

With the inclusion of the correction in eq.
(\ref{eq:1loop-alfatop-alfa}) and of the top contribution from eq.
(\ref{eq:potential-replace}), eq. (\ref{eq:higgsmass}) reads
\begin{align}
   \label{eq:mhiggs}
   m_h^2 & = M_Z^2 \left[1- \frac{3}{4\,{\sqrt{2}}\,{\pi }^2}{G_F} \;
     {{m_t}}^2 \; \log \left(\frac{m^2_Q(ML)
         m_U^2(ML)}{m_t^4}\right)\right] \nonumber \\
   & +4 \sqrt 2 G_F v^4 \delta V_{\textrm{top}}''(v^2;ML).
\end{align}
By means of the relation between $ML$ and $1/L$ as determined from the
minimum equation, $m_h$ is plotted in Fig.  \ref{fig:mhiggs} as
function of $1/L$ only, for three different values of the pole top
mass, $m_t^{pole} = 174.3 \pm 5.1 \textrm{ GeV}$.
\begin{figure}[htbp]
  \centering \includegraphics[width=13cm]{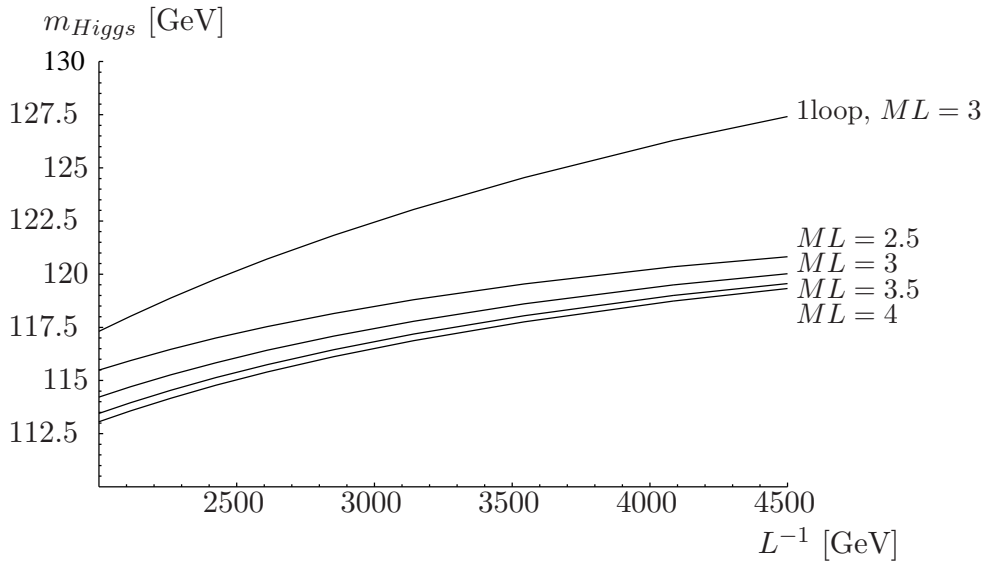}
   \caption{Higgs mass for $m_t^{pole} = 174.3 {\rm \mbox{ }GeV}$ at 
     different fixed values of the top localization parameter. The
     result with the full Higgs potential is compared with the one
     only including the standard top correction at one loop.}
   \label{fig:mhiggs-ML}
\end{figure}
In Figure \ref{fig:mhiggs-ML} we show the band of values that would be
obtained if $ML$ were not related to $1/L$ by the minimum equation
(\ref{eq:pot-minimum}), but kept fixed at values between 2 and 4. This
shows that the precise relation between $ML$ and $1/L$ is almost
irrelevant in order to determine the connection between $m_h$ and the
compactification scale.  In the same Fig. \ref{fig:mhiggs-ML} we also
compare the Higgs mass, calculated on the basis of eq.
(\ref{eq:mhiggs}), with the one that would be obtained from a
minimally improved lowest order formula
\begin{equation}
   \label{eq:mhiggs-naive}
   m_h^2 (\textrm{naive})= M_Z^2 + \frac{3 \sqrt 2}{4 \pi}  G_F m_t^4 
\log \left(\frac{m^2_Q(ML)
   m_U^2(ML)}{m_t^4}\right)
\end{equation}
and $ML= 2 \div 4$. This comparison makes clear that the improved two
loop potential is essential for a better determination of $m_h$.

\subsection{Uncertainties}
\label{sec:uncertainties}

In this way one is naturally lead to the issue of the overall
uncertainties on $m_h$, in particular on Fig. \ref{fig:mhiggs}. We
believe that the deviation from the localized approximation is
properly accounted for by eqs.
(\ref{eq:potential-replace})-(\ref{eq:potential-replace2}). The real
question, then, is the validity of eqs.
(\ref{eq:2loop-quadratic})-(\ref{eq:2loop-quartic}) with an exactly
localized top. To this end, it is necessary to discuss the possible
role of other operators than those in eq. (\ref{eq:lagrangian4D}).
Among those operators, the potentially most important ones are the
kinetic terms of the Higgs and the gauge multiplets localized on the
boundaries. In the notation of eq. (\ref{eq:lagrangian5D}), the
constants $z_H$, $z_a$ in
\begin{equation}
   \label{eq:brane-kinetic-terms}
   \delta {\cal L} =\delta(y) \left[z_H \int {\rm d}^4 \theta
   \, H^{\dagger} e^V H+\sum_a \left(z_a \int {\rm d}^2 \theta
\,   W^{(a)}_{\alpha} W^{(a)}_{\alpha} +{\rm h.c.}\right)\right]
\end{equation}
with $a=SU(3), SU(2),U(1)$ and similar terms localized at $y=L$, have
to be treated as additional parameters.  For them we need an estimate
or a natural assumption for their size.  Since $H$ and
$W^{(a)}_{\alpha}$ are 5D fields, their $z$-factors in eq.
(\ref{eq:brane-kinetic-terms}) have dimension of an inverse mass. Pure
dimensional analysis leads to an estimate $z \sim 1/\Lambda$, where
$\Lambda$ could either be the scale $\Lambda_{np}$ at which some of
the couplings become non perturbative or a cutoff scale
$\Lambda_{cutoff} \gtrsim\Lambda_{np}$ below which our 5D theory
represents the low energy effective description of some more
fundamental theory. By studying the evolution of the Yukawa couplings
of the top and the bottom, localized at $y=0$ and $y=L$ respectively,
one finds that $\Lambda_{np} L =6 \div 10$ at $ML \simeq 3$
\cite{Barbieri:2002uk,Barbieri:2000vh}.  For equal masses, (equal
localization), of the quark hypermultiplets of the third generation,
$M$, it is in fact $\lambda_b$ that becomes non perturbative first. To
produce the physical top and bottom masses, both $\lambda_t$ and $\lambda_b$ depend
on $ML$ and become comparable for $ML=2 \div 2.5$, whereas for higher
$ML$ $\lambda_b$ becomes progressively bigger\footnote{The influence
  of $\lambda_b$ on the Higgs potential is negligible because the
  integrals in $\delta V$ are dominated by low momenta \cite{Barbieri:2002sw}.}.  

Based on these
considerations/assumptions, we take for the dimensionless $Z$-factors
\begin{equation}
   \label{eq:z-factor-estimate}
   Z \equiv \frac{z}{L} \lesssim (10 \div 15) \%.
\end{equation}

On the other hand the introduction of these $z$-factors affects the
radiative Higgs potential or the radiative squark masses $m^2_Q$,
$m_U^2$ only at quadratic order in the dimensionless $Z$'s. On this
basis we expect that the calculations of the Higgs and squark masses
are correct within a few \%. A more critical quantity is $\delta V'$,
since the electroweak and the top contributions cancel quite
accurately against each other and are renormalized by different
factors $(1+{\cal O}(Z^2))$. As already mentioned in Sect.
\ref{sec:Fermi-scale}, although making uncertain the relation between
$ML$ and $1/L$, this has little influence, however, on the relation
between $m_h$ and $1/L$.

\section{Lower values of the compactification scale}
\label{sec:low-1overL}

An interesting question is what happens for $1/L$ lower than $2 \;
{\rm \mbox{ } TeV}$.  Below this value the Higgs mass gets nominally
lower than the experimental bound. Had we drawn the same figure as
Fig. \ref{fig:mhiggs} for lower $1/L$, $m_h$ would have reached values
as low as $105 {\rm \mbox{ }GeV}$ at $1/L \simeq 600 {\rm \mbox{
  }GeV}$ to grow again up to $\simeq 130 {\rm \mbox{ }GeV}$ at $1/L
\simeq 300 {\rm \mbox{ }GeV}$. At the same time, $ML$ progressively
decreases from about 2 at $1/L \simeq 2 {\rm \mbox{ }TeV} $, to zero
at the lowest value of $1/L \simeq 300 {\rm \mbox{ }GeV}$, where one
makes contact with the ``Constrained Standard Model'' of Ref.
\cite{Barbieri:2000vh}. We do not show this plot because in the
intermediate region of $1/L \approx 1 {\rm \mbox{ }TeV} $ or $ML
\simeq 1$, our calculation is not fully reliable: $\delta
V_{\textrm{top}}^{1loop}$ does not clearly dominate over the 2 loop
contribution, which, on the other hand, is only to be trusted for
sufficiently large $ML$.

To make sense of the model at these lower values of $1/L$, one has
also to make sure that the potential with two Higgs doublets, $h$ and
$h_c$, does not get destabilized, given the absence of a bilinear term
$h h^c$. This is possible, without introducing a FI term, by adding a
small common mass, $| M_H L | \lesssim 0.1 $ for the Higgs
hypermultiplets.

\begin{figure}[bhtp]
  \centering \includegraphics[width=13cm]{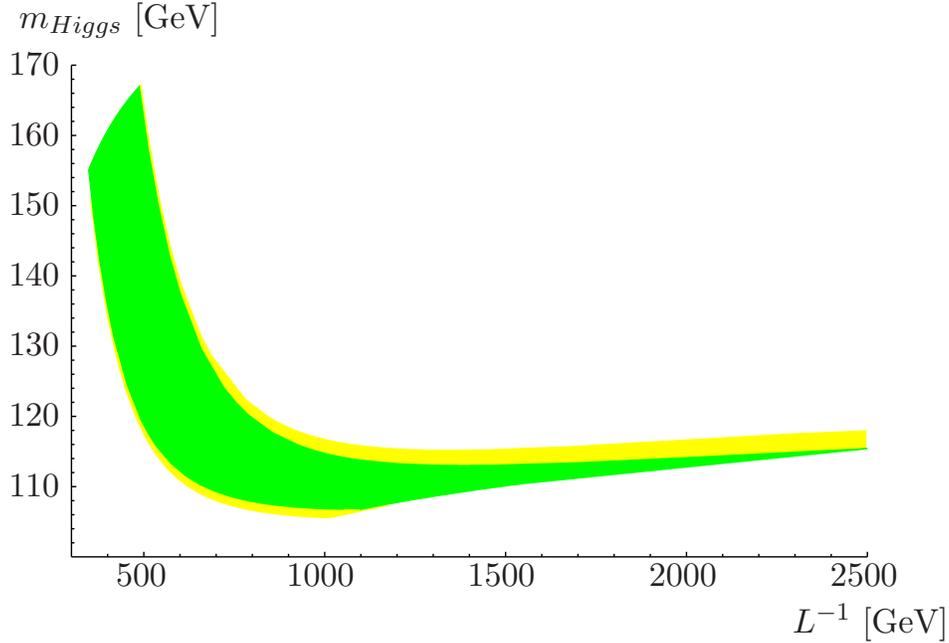}
   \caption{Expected range of Higgs masses with the inclusion of a 
     hypermultiplet Higgs mass and a maximum fine tuning in the slope
     of the potential at $10\%$ level. The darker (lighter) area is
     for $m_t^{\textrm{pole}}=174.3$ ($174.3 \pm 5.1$)}
   \label{fig:mhiggs-MH}
\end{figure}

A non zero $M_H$ does not affect the physical Higgs mass, through
$V''$, but only the determination of the Fermi scale, via $V'$. With
an extra term $(- M_H L)$, present in the right hand side of eq.
(\ref{eq:deltaV-prime}), $1/L$ is not tied anymore to $ML$, which can
in turn vary in a range consistent with a moderate amount of fine
tuning. The result of this is shown in Fig. \ref{fig:mhiggs-MH}.
Different values of $ML$ are used, but always in such a way that no
fine tuning occurs stronger than $10\%$ in the determination of
$G_F^{-1/2}$.  The rise in $m_h$ is due to the stronger influence, for
low $ML$, of $\delta V_{\textrm{top}}^{\textrm{1loop}}$, a fact which
has no correspondence in the MSSM
\cite{Barbieri:2002uk,Barbieri:2000vh}.

Taking into account the uncertainties mentioned above, a value of
$m_h$ marginally consistent with the experimental lower bound of $114
{\rm \mbox{ }GeV}$ cannot be excluded in the entire region of $1/L$.
The existence of independent lower bounds on $1/L$ becomes then of
relevance.  This in turn crucially depends on the masses $M_{Q_i}$,
$M_{L_i}$ for the quark and lepton hypermultiplets of the different
generations, $i=1,2,3$ \cite{Barbieri:2002sw}. Depending on these
masses, modifications of the Fermi constant \cite{Strumia:1999jm}
and/or Flavor Changing Neutral Currents effects \cite{Delgado:1999sv}
are possible by tree level exchanges of KK states of the gauge bosons.
These effects, however, are minimized (made to vanish at tree level)
by the choice $M_{L_i}=0$, $M_{Q_i}=M$. It is in fact possible that
the most important constraint on $1/L$ comes from the electroweak
$\rho$ parameter. Simple dimensional analysis shows that the
correction to the $\rho$ parameter from the towers of top and bottom
states scales as $\delta\rho=c \, v^2/\left(L^2 \Lambda^4\right)$ with
a dimensionless coefficient that could perhaps be as large as 30
\cite{Barbieri:2000vh}.  Although this correction, at fixed $\Lambda L
= 6 \div 10$, vanishes as $L^2$ for small $L$, a value of $1/L$ as
large as $1 {\rm \mbox{ }TeV} $ might be needed to suppress this
$\delta \rho$ correction below the uncertainty of the observed value.

\section{Spectrum}
\label{sec:spectrum}

The most peculiar feature of the spectrum is the heaviness of
gauginos--higgsinos with a mass $\pi / (2 L)$. The lightest
superpartner is therefore a squark or a slepton, most likely charged,
which can be stable or practically stable if a small $U(1)_R$-breaking
coupling is present. The first KK states corresponding to the SM
particles are at $\pi/L$.  The masses of the sfermions of charge $Q$
and hypercharge $Y$ are given by
\begin{equation}
   \label{eq:spectrum}
   m^2=m^2_{tree}+m^2_{rad}+Y m_Z^2-Q m_W^2
\end{equation}
where $m_{tree}$ is the tree level mass, including the Yukawa
contribution, and $m_{rad}$ is the one loop contribution, as in eq.
\ref{eq:mstops}.  As seen in Sect. \ref{sec:quasiloc-top}, both
$m_{tree}$ and $m_{rad}$ depend upon the corresponding localization
parameter $ML$. For $ML=0$, $m_{tree}=\pi / (2 L)$ dominates and the
sfermion becomes degenerate with the gauginos and the higgsinos. For
$ML \gtrsim 1$, $m_{rad}$ dominates and rapidly approaches the
localized limit where
\begin{subequations}
   \label{eq:mrad}
\begin{align}
  m_{rad}(\widetilde Q) &= (370 {\rm \mbox{ }GeV})\frac{1}{L {\rm
      \mbox{ }TeV} } \label{eq:mQ}\\
  m_{rad}(\widetilde U) &= (382 {\rm \mbox{ }GeV})\frac{1}{L {\rm
      \mbox{ }TeV} } \\
  m_{rad}(\widetilde D) &= (310 {\rm \mbox{ }GeV})\frac{1}{L {\rm
      \mbox{ }TeV} } \\
  m_{rad}(\widetilde L) &= (137 {\rm \mbox{ }GeV})\frac{1}{L {\rm
      \mbox{ }TeV} }  \label{eq:mL}\\
  m_{rad}(\widetilde E) &= (79 {\rm \mbox{ }GeV})\frac{1}{L {\rm
      \mbox{ }TeV} }
\end{align}
\end{subequations}
Up to the D-term effects in eq. (\ref{eq:spectrum}), these are lower
values for the sfermion masses.

As we have seen, the localization parameter of the third generation
squarks is correlated with the compactification scale by EWSB. For
$1/L \gtrsim 2 {\rm \mbox{ }TeV} $, eq.  (\ref{eq:mQ}) gives the mass
of the left handed stop and sbottom, split by $m_{tree}(\widetilde
t)=m_t$, whereas $\left(m_{rad} (\widetilde U)^2+m_t^2\right)^{1/2}$
is the mass of the right-handed stop.

For lower values of $1/L$, if a tree level mass of the Higgs
hypermultiplets is present (see Sect. \ref{sec:low-1overL}), the third
generation squarks can be progressively delocalized. In this case eqs.
(\ref{eq:mrad}) give a lower bound, with the overall masses that can
go up to $800 {\rm \mbox{ }GeV}$ even for $1/L$ below $1 {\rm \mbox{
  }TeV} $.

Eqs.  (\ref{eq:spectrum}) and (\ref{eq:mL}) apply also to the masses
of the scalars in the second Higgs hypermultiplet $H_c$, without a vev
and with gauge couplings identical to those of a left-handed slepton
multiplet. Not to undo EWSB, the Higgs hypermultiplets are always
almost fully delocalized, $| M_H L | \leq 0.1$.  Eq. (\ref{eq:mL}) is
strictly valid for $1/L \gtrsim 2 {\rm \mbox{ }TeV}$ and
$m_{rad}(\widetilde H) \gtrsim 270 {\rm \mbox{ } GeV}$.  For smaller
values of $1/L$ a tree level Higgs mass can play a role and can raise
the total mass of the charged and neutral scalars in $H_c$, relative
to eq. (\ref{eq:mL}), by about $100 {\rm \mbox{ }GeV}$ at $1/L \simeq
1 {\rm \mbox{ }TeV} $.

\section{Summary and conclusions}
\label{sec:conclusions}

We have made an accurate calculation of the Higgs potential in a
theory of EWSB triggered by the top Yukawa coupling, where
supersymmetry is broken by boundary conditions on a fifth dimension.
The calculability of the Higgs potential rests on the fact that
supersymmetry breaking is described in terms of a single parameter,
the length $L$ of the compactified extra dimension. This, in turn, is
possible because the zeroth order spectrum has all the extra particles
implied by supersymmetry and/or 5D Poincar\`e invariance at $\pi/(2L)$
or higher.

To define the phenomenology of this proposal, a central issue is the
determination of the range of the compactification scale $1/L$.
Although not with certainty, the ElectroWeak Precision Tests most
likely want a value of $1/L$ above $1 {\rm \mbox{ }TeV} $. This is a
manifestation of the usual ``little hierarchy problem'': the apparent
need of a gap between the scale of physics that triggers EWSB and
$G_F^{-1/2}$ or the Higgs mass, low according to the same EWPT. In the
present case, the relation between $1/L$ and $G_F^{-1/2}$ is
influenced by the level of localization of the top quark and, to a
lesser extent, by a hypermultiplet Higgs mass. These are both
unrenormalized supersymmetric parameters.

The two loop calculation of the Higgs potential shows that $1/L$ can
go up to $4 {\rm \mbox{ }TeV} $ without a significant amount of fine
tuning and with the third generation Yukawa couplings maintaining
perturbativity up to $\Lambda=(6 \div 10)/L$. This is made possible by
the localization of the top near the boundary of the 5th dimension where its Yukawa
coupling is present. In this way the otherwise dominant one-loop top
contribution to the curvature of the potential $V'(v^2)$ gets
exponentially suppressed.  Furthermore the two-loop diagrams involving
the top Yukawa couplings in exactly localized approximation contribute
to $V'(v^2)$ with a negative term that almost cancels the positive
electroweak contribution, so that $L^2 V'(v^2) = \mathcal{O}
(10^{-3})$ for $M L \simeq 2.5 \div 3.5$.  This is the basis for
allowing values of $1/L$ larger than $G_F^{-1/2}$ by about one order
of magnitude.

To make sure that this is at all consistent with the phenomenological
constraints, it is crucial to compare the expected Higgs mass with the
current lower bound. To this purpose, with a reasonable assumption on
the uncertainties induced by boundary $Z$-factors, it has been
necessary to extend to two loop also the calculation of $V''(v^2)$.
For large values of $1/L = 2 \div 4 {\rm \mbox{ }TeV} $, the Higgs
mass is reduced with respect to the naive expectation and is in the
$110 \div 125 {\rm \mbox{ } GeV}$ range. For lower values of $1/L$ the
Higgs mass is more uncertain but can be above the experimental bound
up to the lowest possible values of $1/L$.

The spectrum implied by this picture of supersymmetry and electroweak
symmetry breaking has a few characteristic features. The first KK
states corresponding to the SM particles are at $\pi/L$. Also all
gauginos and higgsinos are heavy, with a mass at $\pi/(2 L)$.
Therefore the lightest superpartner is a sfermion, most likely
charged, which can be stable or practically stable for collider
experiments \cite{Barbieri:2002sw,Barbieri:2000vh}. Finally one also
expects relatively light scalars, one charged and one neutral,
belonging to the $H_c$ hypermultiplet, introduced to cancel the FI
term. They are coupled to the gauge bosons as a lepton doublet but
have unknown couplings to quarks and leptons.

\section*{Acknowledments}
\label{sec:acknowledge}

We thank P.~Slavich for useful comments concerning the two loop
corrections to the Higgs mass in the MSSM. This work has been
partially supported by MIUR and by the EU under TMR contract
HPRN-CT-2000-00148.

\appendix

\section{The 2 loop contribution to the Higgs potential}

\label{app:2loop}

In this appendix we give some details of the 2 loop contribution to
the Higgs potential $\delta V_{\textrm{top}}^{\textrm{2 loop}}$.

At the 2-loop level there are two contributions.  The first one comes
from the expansion, in eq.  (\ref{eq:1loop-top-localized}), of the one
loop corrections in eqs. (\ref{eq:renormalizations}) after expressing
$m_{0,t}$, $m_{0,Q}$ and $m_{0,U}$ in terms of the renormalized masses $m_t$, $m_Q$ and $m_U$
respectively. The second contribution is a pure 2-loop correction and
corresponds to the diagrams of Fig. \ref{fig:2ndtopology} in terms of
the top superfields $U,Q$, the Higgs superfield $H$ and the $SU(3)$
vector superfiled $V$.  In localized approximation for $U$ and $Q$, only $V$ and
$H$ are 5-dimensional superfields.

Defining
\begin{equation}
C(x)=x \coth (x) \hspace{2cm} T(x)=x \tanh (x) \label{eq:def-CT}
\end{equation}

the pure 2-loop gauge correction (arising from the diagrams
(\ref{fig:2ndtopology})-b) is given by the following expression

\begin{align*}
  V_{\textrm{2loop,gauge}}=4\,{{g_s}}^2\, \int & \frac{d^4 p}{(2
    \pi)^4} \frac{d^4 q}{(2 \pi)^4} \frac{T(q
    L)}{q^2} \times \\
  \times & \left[ \frac{-2\, q^2} {\left( {\left( p - q \right) }^2 +
        {{m_Q}}^2 \right) \, \left( p^2 + {{m_t}}^2 \right) } -
    \frac{2\,p^2} {\,\left( {\left( p - q \right) }^2 + {{m_Q}}^2
      \right) \,
      \left( p^2 + {{m_t}}^2 \right) } \right. \\
  & + \frac{2\,{\left( p - q \right) }^2} {\,\left( {\left( p - q
        \right) }^2 + {{m_Q}}^2 \right) \, \left( p^2 + {{m_t}}^2
    \right) } - \frac{2\, q^2} {\left( p^2 + {{m_t}}^2 \right) \,
    \left( {\left( p - q \right) }^2 +
      {{m_U}}^2 \right) } 
\end{align*}
\begin{align*}
\hspace{2cm}
  & - \left. \frac{2\,p^2} {\,\left( p^2 + {{m_t}}^2 \right) \, \left(
        {\left( p - q \right) }^2 + {{m_U}}^2 \right) } +
    \frac{2\,{\left( p - q \right) }^2} {\,\left( p^2 + {{m_t}}^2
      \right) \, \left( {\left( p - q \right) }^2 +
        {{m_U}}^2 \right) }  \right] \\
  +4\,{{g_s}}^2\, \int & \frac{d^4 p}{(2 \pi)^4} \frac{d^4 q}{(2
    \pi)^4} \frac{C(q L)}{q^2} \times \\
  \times & \left[ \frac{4} {\,\left( {\left( p - q \right) }^2 +
        {{m_Q}}^2 \right) } + \frac{q^2} {\left( p^2 + {{m_Q}}^2
      \right) \, \left( {\left( p - q \right) }^2 + {{m_Q}}^2 \right)
    } \right. \\ & - \frac{p^2} {\,\left( p^2 + {{m_Q}}^2 \right) \,
    \left( {\left( p - q \right) }^2 + {{m_Q}}^2 \right) } -
  \frac{{\left( p - q \right) }^2} {\,\left( p^2 + {{m_Q}}^2 \right)
    \, \left( {\left( p - q \right) }^2 +
      {{m_Q}}^2 \right) }\\
  & + \frac{2 \, q^2} {\left( p^2 + {{m_t}}^2 \right) \, \left(
      {\left( p - q \right) }^2 + {{m_t}}^2 \right) } - \frac{2\,p^2}
  {\,\left( p^2 + {{m_t}}^2 \right) \, \left( {\left( p - q \right)
      }^2 +
      {{m_t}}^2 \right) }\\
  & - \frac{2\,{\left( p - q \right) }^2} {\,\left( p^2 + {{m_t}}^2
    \right) \, \left( {\left( p - q \right) }^2 + {{m_t}}^2 \right) }
  - \frac{8\,{{m_t}}^2} {\,\left( p^2 + {{m_t}}^2 \right) \, \left(
      {\left( p - q \right) }^2 +
      {{m_t}}^2 \right) }\\
  & + \frac{4} {\,\left( p^2 + {{m_U}}^2 \right) } + \frac{q^2}
  {\left( p^2 + {{m_U}}^2 \right) \, \left( {\left( p - q \right) }^2
      +
      {{m_U}}^2 \right) }\\
  & -\left. \frac{p^2} {\,\left( p^2 + {{m_U}}^2 \right) \, \left(
        {\left( p - q \right) }^2 + {{m_U}}^2 \right) } -
    \frac{{\left( p - q \right) }^2} {\,\left( p^2 + {{m_U}}^2 \right)
      \, \left( {\left( p - q \right) }^2 + {{m_U}}^2 \right) }
  \right]
\end{align*}

Analogously the pure 2-loop Yukawa correction (arising from the diagrams
(\ref{fig:2ndtopology})-a) is given by the following expression

\begin{align*}
  V_{\textrm{2loop,Yuk.}}= 3\,{{y_t}}^2 \int & \frac{d^4 p}{(2 \pi)^4}
  \frac{d^4 q}{(2 \pi)^4} \frac{T(q L)}{q^2} \times \\ \times & \left[
    \frac{{\left( p - q \right) }^2} {\left( {\left( p - q \right) }^2
        + {{m_Q}}^2 \right) \, \left( p^2 + {{m_t}}^2 \right) } -
    \frac{p^2} {\left( {\left( p - q \right) }^2 + {{m_Q}}^2 \right)
      \, \left( p^2 + {{m_t}}^2 \right) } \right.  \\ & - \frac{q^2}
  {\left( {\left( p - q \right) }^2 + {{m_Q}}^2 \right) \, \left( p^2
      + {{m_t}}^2 \right) } - \frac{p^2} {\left( {\left( p - q \right)
      }^2 + {{m_Q}}^2 - {{m_t}}^2 \right) \, \left( p^2 + {{m_t}}^2
    \right) }\\ & + \frac{{\left( p - q \right) }^2} {\left( {\left( p
          - q \right) }^2 + {{m_Q}}^2 - {{m_t}}^2 \right) \, \left(
      p^2 + {{m_t}}^2 \right) } - \frac{q^2} {\left( {\left( p - q
        \right) }^2 + {{m_Q}}^2 - {{m_t}}^2 \right) \, \left( p^2 +
      {{m_t}}^2 \right) }\\ & - \frac{1}{p^2 + {{m_U}}^2} + \frac{p^2}
  {{\left( p - q \right) }^2\, \left( p^2 + {{m_U}}^2 \right) } -
  \frac{q^2} {{\left( p - q \right) }^2\, \left( p^2 + {{m_U}}^2
    \right) }\\ & + \frac{p^2} {\left( {\left( p - q \right) }^2 +
      {{m_t}}^2 \right) \, \left( p^2 + {{m_U}}^2 \right) } -
  \frac{{\left( p - q \right) }^2} {\left( {\left( p - q \right) }^2 +
      {{m_t}}^2 \right) \, \left( p^2 + {{m_U}}^2 \right) }\\ & -
  \left. \frac{q^2} {\left( {\left( p - q \right) }^2 + {{m_t}}^2
      \right) \,
      \left( p^2 + {{m_U}}^2 \right) } \right] 
\end{align*}
\begin{align*}
\hspace{1.5cm}
  + 3\, {{y_t}}^2 \int & \frac{d^4 p}{(2 \pi)^4} \frac{d^4 q}{(2
    \pi)^4}
  \frac{C(q L)}{q^2} \times \\
  \times & \left[ \frac{p^2} {\left( p^2 + {{m_Q}}^2 \right) \, \left(
        {\left( p - q \right) }^2 + {{m_Q}}^2 \right) } +
    \frac{{{m_Q}}^2 - {{m_t}}^2} {\left( p^2 + {{m_Q}}^2 \right) \,
      \left( {\left( p - q \right) }^2 + {{m_Q}}^2 \right) } \right. \\
  & + \frac{p^2} {\left( p^2 + {{m_Q}}^2 \right) \, \left( {\left( p -
          q \right) }^2 + {{m_Q}}^2 - {{m_t}}^2 \right) } +
  \frac{{{m_Q}}^2 - {{m_t}}^2} {\left( p^2 + {{m_Q}}^2 \right) \,
    \left( {\left( p - q \right) }^2 + {{m_Q}}^2 - {{m_t}}^2 \right) }
  \\ & - \frac{1}{p^2 + {{m_t}}^2} - \frac{p^2} {{\left( p - q \right)
    }^2\, \left( p^2 + {{m_t}}^2 \right) } + \frac{q^2} {{\left( p - q
      \right) }^2\, \left( p^2 + {{m_t}}^2 \right) }\\ & - \frac{p^2}
  {\left( p^2 + {{m_t}}^2 \right) \, \left( {\left( p - q \right) }^2
      + {{m_t}}^2 \right) } - \frac{{\left( p - q \right) }^2} {\left(
      p^2 + {{m_t}}^2 \right) \, \left( {\left( p - q \right) }^2 +
      {{m_t}}^2 \right) }\\ & + \frac{q^2} {\left( p^2 + {{m_t}}^2
    \right) \, \left( {\left( p - q \right) }^2 + {{m_t}}^2 \right) }
  + \frac{1}{p^2 + {{m_U}}^2}\\ & + \frac{q^2} {\left( {\left( p - q
        \right) }^2 + {{m_Q}}^2 \right) \, \left( p^2 + {{m_U}}^2
    \right) } + \frac{q^2} {\left( {\left( p - q \right) }^2 +
      {{m_Q}}^2 - {{m_t}}^2 \right) \, \left( p^2 + {{m_U}}^2 \right)
  }\\ & + \left. \frac{{\left( p - q \right) }^2} {\left( p^2 +
        {{m_U}}^2 \right) \, \left( {\left( p - q \right) }^2 +
        {{m_U}}^2 \right) } + \frac{-{{m_t}}^2 + {{m_U}}^2} {\left(
        p^2 + {{m_U}}^2 \right) \, \left( {\left( p - q \right) }^2 +
        {{m_U}}^2 \right) } \right] \,
\end{align*}

In the 2-loop potential, we have used the physical (renormalized)
quantities $m_t^2, \, m_U^2, \, m_Q^2$
because the corrections are of higher order. When one takes the
derivatives of the potential one has to remember that
\begin{equation}
v^2 \frac{\textrm{d}}{\textrm{d} v^2} = m_t^2 \left( 
\frac{\partial}{\partial m_t^2} +  \frac{\partial}{\partial m_U^2} + 
\frac{\partial}{\partial m^2_Q} \right)
\end{equation}
because all the 3 masses depend on the VEV $v$.

The integrals in $p$ can be performed analitically. Then, to get the
leading logarithmic contributions as $L \rightarrow 0$, one can use
the asymptotic behaviour of the $q$-integrand functions. In this way
one gets the results given in eqs.
(\ref{eq:2loop-quadratic})-(\ref{eq:2loop-quartic}).

\section{1-loop Renormalization functions at order $\alpha_t$, $\alpha_s$}
\label{app:b-factors}

In order to compute the physical masses in Eqs.
(\ref{eq:renormalization-conditions-top})-(\ref{eq:renormalization-conditions-stopu}),
the one loop corrections ${\cal O}(\alpha_t,\alpha_s)$ to the
propagators of the $U$,$Q$-multiplets and to the Yukawa vertex are
needed.

At order $\alpha_t$ the propagators of the top, stop and auxiliary
field get corrected from the exchange of the Higgs supermultiplet H
and a U (or Q) quark multiplet.  

These corrections can be parameterized as usual
\nopagebreak

\includegraphics{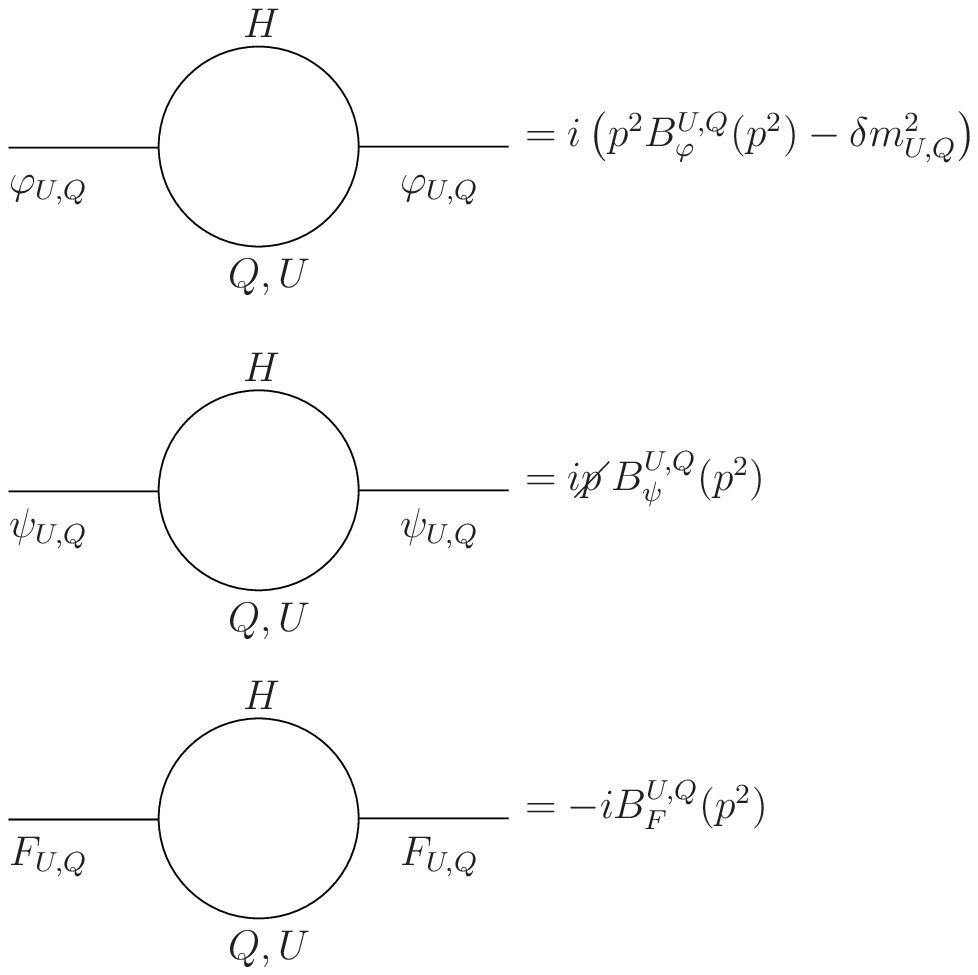}

where we have used a superfield notation in the loop and an Euclidean
external momentum $p$.  The Yukawa vertex receives no direct
correction at order $\alpha_t$ because of the non-renormalization
properties of the superpotential.

All the quantities defined above, to the order $y_t^2$, are given by
\begin{subequations}
\begin{align}
  B_F^U (0) &= - y_t^2 \int \frac{d^4 q}{(2 \pi)^4} C \left(q L
  \right) \left[ \frac{1}{q^2+m^2_{Q}+m_{t}^2}+
    \frac{1}{q^2+m^2_{Q}} \right] \\
  B_F^Q (0) &= - y_t^2 \int \frac{d^4 q}{(2 \pi)^4} C \left( q L
  \right) \left[ \frac{1}{q^2+m^2_{Q}+m_{t}^2} \right]\\
  B_{\varphi}^Q (0) &= y_t^2 \int \frac{d^4 q}{(2 \pi)^4}
  \frac{1}{q^2} \left\{ C\left( q L \right)
    \left[\frac{q^2(m_{t}^2+m^2_{U})}{(q^2+m_{t}^2+m^2_{U})^3}
      -\frac{m_{t}^2(m_{t}^2+m^2_{Q})}{(q^2+m_{t}^2+m^2_{Q})^3}
    \right]
  \right. \nonumber \\
  & \hspace{2.5cm} \left. - T\left( q L \right)
    \left[\frac{q^4+3m_{t}^2 q^2}{(q^2+m_{t}^2)^3}\right]\right\} \\
  B_{\varphi}^U (0) &=y_t^2 \int \frac{d^4 q}{(2 \pi)^4} \frac{1}{q^2}
  \left\{ C\left( q L \right)
    \left[\frac{q^2(m_{t}^2+m^2_{Q})}{(q^2+m_{t}^2+m^2_{Q})^3}
      -\frac{m_{t}^2(m_{t}^2+m^2_{U})}{(q^2+m_{t}^2+m^2_{U})^3}
      +\frac{q^2 m^2_{Q}}{(q^2+m^2_{Q})^3} \right] \right. \nonumber \\
  & \hspace{2.5cm} \left. - T\left( q L \right)
    \left[\frac{q^4+3m_{t}^2 q^2}{(q^2+m_{t}^2)^3}+\frac{1}{q^2}
    \right]\right\}\\
  B_{\psi}^Q (0) &=- \frac{y_t^2}{2} \int \frac{d^4 q}{(2 \pi)^4}
  \frac{1}{q^2} \left\{ C\left( q L \right)
    \frac{q^2+2m_{t}^2}{(q^2+m_{t}^2)^2} + T\left( q L \right)
    \frac{q^2}{(q^2+m^2_{U}+m_{t}^2)^2}\right\} 
\end{align}
\begin{align}
  B_{\psi}^U (0) &=-\frac{y_t^2}{2} \int \frac{d^4 q}{(2 \pi)^4}
  \frac{1}{q^2} \left\{ C\left( q L \right) \left[
      \frac{q^2+2m_{t}^2}{(q^2+m_{t}^2)^2}+\frac{1}{q^2}
    \right] \right. \nonumber \\
  & \hspace{2.5cm} \left. - T\left( q L \right) \left[
      \frac{q^2}{(q^2+m^2_{Q}+m_{t}^2)^2}
      +\frac{q^2}{(q^2+m^2_{Q})^2} \right]\right\} \\
  \delta m^2_{Q} &=y_t^2 \int \frac{d^4 q}{(2 \pi)^4} \frac{1}{q^2}
  \left\{ C\left( q L \right)
    \left[\frac{q^2+m^2_{Q}}{q^2+m^2_{Q}+m_{t}^2}+\frac{q^2}{q^2+m_{
          0,u}^2+m_{t}^2}\right] - T\left( q L \right) \frac{2
      q^2}{q^2+m_{t}^2} \right\}\\
  \delta m^2_{U} &=y_t^2 \int \frac{d^4 q}{(2 \pi)^4}\frac{1}{q^2}
  \left\{ C\left( q L \right)
    \left[\frac{q^2+m^2_{U}}{q^2+m^2_{U}+m_{t}^2}+1+\frac{q^2}{q^2+m
        _{q}^2+m_{t}^2}+\frac{q^2}{q^2+m^2_{Q}}\right] \right.
  \nonumber \\
  & \hspace{2.5cm} \left. - T\left( q L \right)\left[ \frac{2
        q^2}{q^2+m_{t}^2}+2\right]\right\}
\end{align}
\end{subequations}
where the functions $C\left(x\right)$ and $T\left(x\right)$ are given
in (\ref{eq:def-CT}). The integration over the
momentum $q$ has to be performed on Euclidean space.

At order $\alpha_s$ only the propagators of the top and the stops, but
not of their auxiliary fields, get corrected from the exchange of the
$SU(3)$ gauge supermultiplet $V$ and of a quark multiplet. Performing
the calculation in the Wess-Zumino gauge there is also a direct
correction to the Yukawa interaction, so that
\begin{figure}[!ht]
  \includegraphics{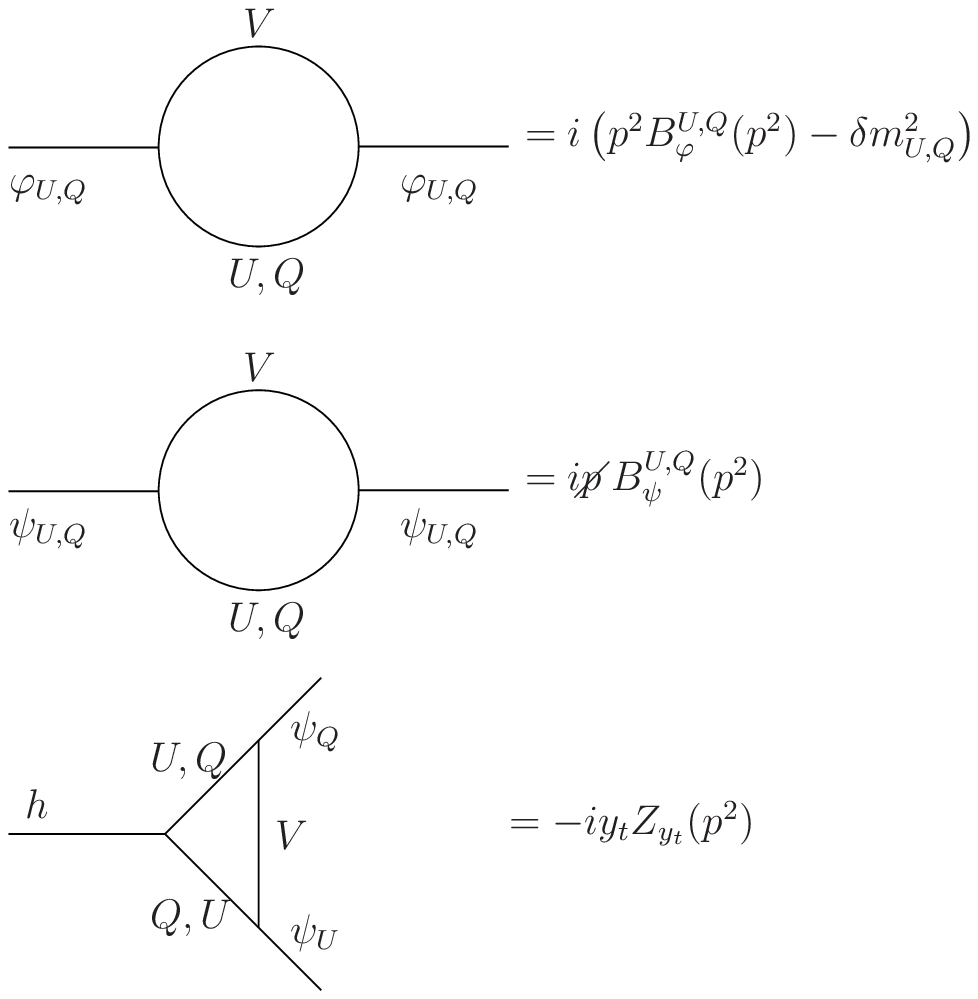}
\end{figure}

Parameterizing these $g_s^2$-corrections as in the $y_t^2$ case, one
has
\begin{subequations}
\begin{align}
  B_{\varphi}^i (0) &= \frac{8}{3} g_s^2 \int \frac{d^4 q}{(2 \pi)^4}
  \frac{1}{q^2} \left\{ C\left( q L \right) \frac{q^2+2(
      m_{i}^2+m_{t}^2)}{(q^2+m_{i}^2+m_{t}^2)^2}- T\left(q
      L\right) \frac{q^4+3 q^2 m_{t}^2}{(q^2+m_{t}^2)^3} \right\}\\
  B_{\psi}^i (0) &= -\frac{4}{3} g_s^2 \int \frac{d^4 q}{(2 \pi)^4}
  \frac{1}{q^2} \left\{ C\left( q L \right)
    \frac{q^2+2m_{t}^2}{(q^2+m_{t}^2)^2} + T\left(q L\right)
    \frac{q^2}{(q^2+m_{i}^2+m_{t}^2)^2} \right\}\\
  \delta m^2_{i} &= \frac{16}{3} g_s^2 \int \frac{d^4 q}{(2 \pi)^4}
  \frac{1}{q^2} \left\{ C\left( q L \right) - T\left(q L\right)
    \frac{q^2}{q^2+m_{t}^2} \right\}\\
  Z_{y_t} (0) & =\frac{16}{3} g_s^2 \int \frac{d^4 q}{(2 \pi)^4}
  \frac{1}{q^2} C\left( q L \right) \frac{q^2}{(q^2+m_{t}^2)^2}
\end{align}
\end{subequations}
where $i=U,Q$.

All the quantities defined above are regular in the IR except
$B_{\psi}^U$.  Because we are interested only in the logarithmic
contributions to $\delta V''(v^2)$, we can evaluate the $B_{\varphi}$
and $B_F$ functions at vanishing external momentum.  Instead, the
functions $B_{\psi}$ and $Z_{y_t}$, involved in eq.
(\ref{eq:renormalization-conditions-top}), have to be evaluated at
$p^2=m_t^2$.  Given their expressions at $p^2=0$ one has to add
\begin{align}
  B_{\psi}^Q \left(m_t^2 \right) - B_{\psi}^Q \left(0 \right) &=
  \frac{\alpha_s}{6 \pi} + \frac{\alpha_t}{16 \pi}\\
  B_{\psi}^U \left(m_t^2 \right) - B_{\psi}^U \left(0 \right) &=
  \frac{\alpha_s}{6 \pi} + \frac{\alpha_t}{16 \pi} + 4 \pi \alpha_t
  \int \frac{d^4 q}{(2 \pi)^4} \left[ \right.  \frac{1}{2\,q^4} +
  \frac{4\, \log (\frac{2\,q} {{m_t} + {\sqrt{4\,q^2 +
          {{m_t}}^2}}})}{{m_t}\,{\left( 4\,q^2 + {{m_t}}^2
      \right) }^{\frac{3}{2}}} \nonumber\\
  & \hspace{4.7cm} - \frac{1}{q^2\, \left( 4\,q^2 + {{m_t}}^2 \right)
  }
  \left. \right] \label{eq:shift-Bu}\\
  Z_{y_t} \left(m_t^2 \right) - Z_{y_t} \left(0 \right) &= \frac{4
    \alpha_s}{3 \pi} \left(2-3 \log 2 \right)
\end{align}

Note that the IR divergence in (\ref{eq:shift-Bu}) cancels the one in
$B_{\psi}^U(0)$.

\end{document}